# FROM PILES OF GRAINS TO PILES OF VORTICES

# DE LAS PILAS DE GRANOS A LAS PILAS DE VÓRTICES

E. Altshuler[1]

a) [1]Grupo de Sistemas Complejos y Física Estadística, Facultad de Física, Universidad de La Habana, 10400 Habana, Cuba E-mail: ealtshuler@fisica.uh.cu†

†autor para la correspondencia

**Abstract.** One of the most thrilling features of Physics is the possibility to establish analogies between apparently distant areas. Here we explain the parallel between a pile of grains interacting mechanically with each other, and a "pile" of superconducting vortices. In both cases the macroscopic slope of the pile is maintained by a very nonlinear avalanche process. Furthermore, both types of piles logarithmically relax in time due to "agitational" or thermal effects, aiming at a state of equilibrium.

**Resumen.** Uno de los aspectos más apasionantes de la Física es la posibilidad de establecer analogías entre áreas aparentemente distantes entre sí. Aquí explicamos el paralelo entre una pila de granos que interactúan mecánicamente entre sí, y una "pila" de vórtices superconductores. En ambos casos la pendiente de la pila se mantiene mediante avalanchas altamente no lineales. Además, ambos tipos de pilas se relajan en el tiempo debido a efectos "agitativos" o térmicos, apuntando hacia un estado de equilibrio.

It was perhaps Galileo who first realized that provocative titles are crucial to capture the attention of potential readers of science books. In *Dialogues concerning two new sciences* (1638) [1], he actually introduces two new disciplines: I might call them *the science of motion*, and *materials science*. The book is an undisputed masterpiece, so it is hard to criticize its somewhat pompous title. Curiously, the idea of titles announcing new sciences has become fashionable again these days.

Judge by yourself: "Sync: The Emerging Science of Spontaneous Order (2004) [2], "Nexus: Small Worlds and the Groundbreaking Science of Networks" (2003) [3], "A New Kind of Science" (2002) [4] and "How Nature Works: The Science of Self-Organized Criticality" (1996) [5]. The latter, written by the Danish physicist Per Bak, immediately attracted many followers as well as detractors. While there is consensus today that the book's title is not a paradigm of objectivity, there is no doubt that the subject of Self-Organized Criticality (SOC) concentrated a lot of attention on granular matter by cleverly using the granular pile as a physical paradigm.

In any case, I believe that SOC strongly contributed to the incorporation of physicist to the study of sandpiles (and, in general, granular matter) [6-10], a subject almost exclusively tackled by engineers before the end of the XX century (See, for example, [11]).

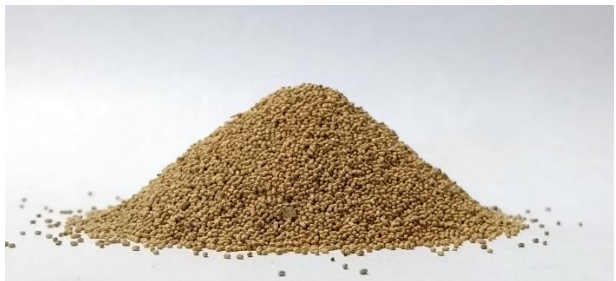

**Figure 1.** A pile of termite excrement. The width of the base of the actual pile is somewhat larger than 5 cm.

However, you do not need Self-Organized Criticality to realize that the granular pile shown in Fig. 1 is extraordinary. And it is not because you rarely see a beautiful picture of a pile of drywood termite excrement: it is extraordinary just for being a pile. First of all, we must notice that most of its slope is straight, in spite of the fact that it extends for a length much larger than the size of the individual grains. How is it possible that grains, which only seem to interact when they are directly touching their next neighbors, can organize themselves in such a large structure? Human beings can form almost perfect lines and nice platoons, but it is generally achieved by means of some kind of centralized organization that directly reaches many individuals at various distances from the center. But there is no such thing in a granular pile. In addition, if you form the pile by slowly adding grains from the top, its smooth slope results from landslides (or avalanches). You have also probably observed that the general shape of the pile is very robust: if you form a new pile on a table by dropping grains of the same kind from above you will reach approximately the same angle relative to the horizontal, even if you do not deliver them exactly in the same way. You will get the same robustness using beads, sand, termite excrement... It illustrates *self-organization*.

SOC proposes a mechanism where local interactions between neighboring grains are able to explain the emergence of a pile with straight slopes. The simplest computational way to work out the idea is known as the BTW model [12], illustrated in Fig. 2. We have a table on which we pour identical grains at random places, one by one. Then, we apply very simple rules trying to mimic the way a "real" pile becomes locally unstable. In the digital world of the model, the grains cannot land everywhere, but just on 9 sites of a grid, arranged as illustrated in the upper row of Fig. 2. If one site reaches a threshold of 4 grains, it is emptied out: each of the four grains moves to the neighboring sites located at the top, bottom, right and left relative to the initial one –that is called a *toppling event*.

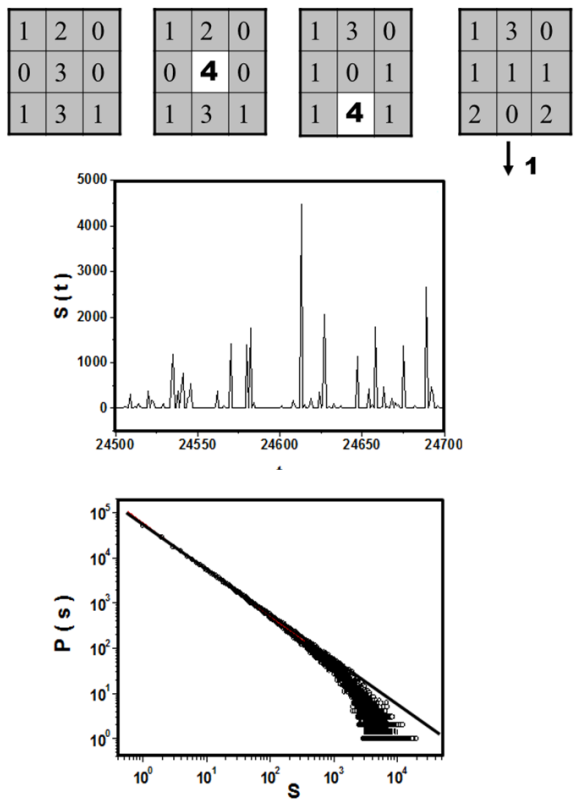

Figure 2. The BTW model in images. Upper row: evolution of an avalanche in a small system of 9 cells. Middle panel: Temporal evolution of avalanches for a much bigger system. Bottom panel: Statistical distribution of avalanche sizes.

If, as a consequence of this action, any of these four sites reaches the threshold of 4 grains, the process repeats again, until all sites within the 9-site grid contain less than 4 grains. In the process, some grains may eventually abandon the 9-site region through its boundary. Only when all toppling events are completed, a new grain is added to the system from the top. Adding grains is a patient and slow process; avalanches are unexpected and fast: the system is very *nonlinear.*

To illustrate these rules, the first graph at the left of the top row in Fig. 2 shows a "quiet moment" in the grid, where all cells contain less than 4 grains (the numbers correspond to the number of grains on each site). In the second graph, one grain has been added to the center cell, which makes it reach the 4-grain threshold. So, the next graph shows how the center cell has been emptied, because all grains have emigrated from it to the four neighboring sites, which accordingly have increased in 1 their number of grains. In the process, the bottom center cell has reached the 4-grain threshold, so the last graph indicates that it has been subsequently emptied, and additional grains have been added to the cells located above it, and to its right and left… a fourth grain had nowhere to go, and was removed through the bottom boundary of the table (If one averages the number of grains added to the grid from top over a long period, it must be equal to the average number of grains abandoning the grid through its border. But it does not mean that you have always one grain exiting when one grain is added: sometimes, many grains can be added and no grain abandons the system… and eventually you may add one grain, and a bunch of grains abandon the system).

Finally, the system is back in calm –all cells are below the threshold– so a new grain can be added from top. If we define an avalanche as the number of toppling events between one addition and the next, we have described here an avalanche of value 2 (you can define many types of avalanches, like the number of sites involved in toppling events and the number of grains that abandoned the system, called off-the-edge avalanches). Of course, we cannot go far with this 9-cell table: the middle and bottom graphs in Fig. 2 illustrate results from a much larger system, studied for a much longer time. The middle graph shows the temporal evolution of avalanche sizes: the horizontal axis corresponds to the time, which is equivalent to the number of grains added from top in the BTW's model; while the vertical axis corresponds to the avalanche size that we will call *s*. So, for example, we can identify an avalanche involving 2000 toppling events taking place when the grain number 24625 was added to the system. Looking at the whole graph, it seems that there are avalanches of many different sizes: no specific size seems to dominate. In fact, there could be avalanches involving the whole grid: somehow, short-range interactions between neighboring grains can involve the whole pile.

Let us now sort the avalanches by size for a very long experiment: for example, we count 100 000 avalanches of size 1, 10 000 avalanches of size 10, 1000 avalanches of size 100, and so on. Then, we plot the results as shown in the bottom graph of Fig. 2 (open circles), that we will call an avalanche size distribution (ASD). The fact that the plot decreases, indicates that large avalanches are rare, while small avalanches are common. Moreover, the data follows a power law (indicated as a solid line): $P(s) \propto s^{-\alpha}$. This kind of distribution indicates that, if you look at the bottom left graph within any time window, the graph looks exactly the same: it is "fractal" in the time domain. For the particular case of the BTW model, the power law has $\alpha = 1$.

Another fundamental (and often forgotten) result from the BTW model is that it gives piles with nearly straight slopes, in spite of the fact that the toppling rules are "local". So, this simple computational automaton tries to justify why grains that only interact with their next neighbors can produce macroscopic structures much larger than the grains themselves. Moreover, small changes in the rules of the BTW model still produce piles with straight slopes and power-law distributed avalanches. If, for a given set of the rules, the resulting power law shows a robust slope value, statistical physicists say they have discovered a new "universality class" (In real experiments, the value of $\alpha$ may move within the range between 1 and 2). In other words, our slowly-fed digital piles are very robust in terms of shape, and that robustness is connected with a "slope-adjusting mechanism" mediated by power-law distributed avalanches. The robustness can be connected to self-organization, and the power-law distribution reminds critical phenomena– an older and very successful field of Physics [13]. So Bak and co-workers called their idea "Self-Organized Criticality". It is safe to say that no one minds too much the self-organization ingredient of the theory, but the Criticality part has been widely criticized. And for good reason: many early attempts to find power-law distributed avalanches in real piles did not show undisputable power laws.

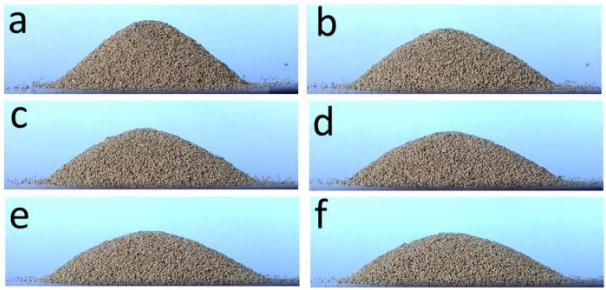

Figure 3. Relaxation of a pile of excrement. (a) Original pile. (b), (c), (d), (e) and (f) are images of the piles after 1, 2, 3, 4 and 5 taps, respectively.

Some of those early experiments were made is a rush inspired by the BTW model, and used various avalanche definitions. It was not clear if the comparison between the resulting avalanche distributions and those generated by BTW-like models were rigorous enough. For example, one of those experiments consisted in slowly pouring sand on the circular plate of a very precise digital scale, and measuring the variations in the pile's mass. Those avalanches were only related to grains falling off the plate: They mimicked best "off the edge" events like the one illustrated in the top right picture of Fig. 2. In fact, one cannot expect a true power law for the statistical distribution of those avalanches. Other experiments involving avalanche definitions closer to the BTW model were performed on piles that were too small to produce trustable statistical distributions of avalanche sizes: the existence or not of a power law could only be assessed by careful manipulations of experimental data related to finite size scaling.

However, carefully controlled experiments in bi-dimensional piles of beads have shown that the avalanches are indeed distributed following a nice power-law, with a slope $\alpha = 1.6$ [14,15]. As an extra benefit, being the slope larger than 1, it can be argued that there exists the possibility of predicting the occurrence of a large avalanche based on the previous history of avalanches –a controversial idea that might be applied to earthquakes, another phenomenon consisting in "tectonic plate slippages" distributed as a power law [16].

The BTW model proposes an avalanche-like mechanism by which a granular pile reaches a meta-stable state with nice straight slopes as it is slowly fed with new grains. However, it does not deal with the problem of how a static pile reaches the true equilibrium. In order to make the pile relax into a stable equilibrium state, we may apply the method shown in Fig. 3. There, panel (a) shows a picture of a termite excrement pile formed by slowly adding grains from top. The pile has been formed on the surface of a cardboard box, which visibly vibrates as it is finger-tapped. The second, third… sixth pictures shown in Fig. 3 correspond to the original pile after applying one, two… five taps on the cardboard. Notice that the slope of the pile decreases very slowly. Common sense indicates that, after an infinite number of taps, the pile would become completely flat, reaching a state of stable equilibrium. When a rigorous experiment is performed, it can be shown that the slope of the pile decreases logarithmically with the number of taps [17].

One of the most thrilling possibilities in science is establishing analogies between fields that are apparently very far away from each other. Here we have one of those beautiful examples: as in the case of the tapped granular pile, magnetization also relaxes logarithmically in the case of superconductors as time goes by. Type II superconductors are peculiar materials that do not allow magnetic fields to penetrate them… up to a certain point. If the magnetic field reaches the so-called "first critical field", a compromise is established between the magnetic field that "wants to penetrate the material" and the material, who "doesn't want to be penetrated": the field gets in, but not flooding it as a continuum. It enters as tiny field-containing spots surrounded by current swirls called Abrikosov vortices [18]. Two forces compete in the penetration process. On the one hand, vortices repel each other at short distances –just as grains of sand repel each other when they collide. Thanks to that, as the external field eagerly "pushes" the vortices in through the materials' boundary, the vortices push each other inside –just as the gravitational force make grains move down the slope of a pile. On the other hand, non-superconductive defects in the superconductor (called "pinning centers") tend to "trap" the vortices, acting as a barrier against their penetration –in the case of grains, they are locally trapped into little bumps on the surface of the pile, preventing them to roll all the way down. The competition between the two effects results in a "vortex pile" where the vortex density is large near the borders of the materials, and decays towards the center, forming a straight slope. It is analogous to slowly filling a shoebox with grains of sand from its borders [19,20].

If we stop adding grains and the shoebox is shaken, the granular slopes will relax inside the box. The same happens with our microscopic vortices, but the role of shaking is played by the agitation associated to conventional temperature: the vortex pile slope will decrease slowly in time, resulting in a logarithmic relaxation of the magnetization –which is just a macroscopic way to measure it. Curiously, this effect –called *flux creep*– was studied in detail years before granular relaxation for conventional superconductors [21], and then for high temperature superconductors [22-24], in spite of the fact that it requires the use of relatively sophisticated equipment and cooling systems.

In principle, one could modify the rules of the BTW model in order to understand relaxation, but now I prefer good old-fashioned statistical physics. The top left panel of Fig. 4 sketches the slope of a granular pile at the very moment the tapping starts. At the right, I show the corresponding "washboard" potential landscape, where grains are trapped into potential wells, which mimics the trapping of grains at the irregularities of the pile's surface.

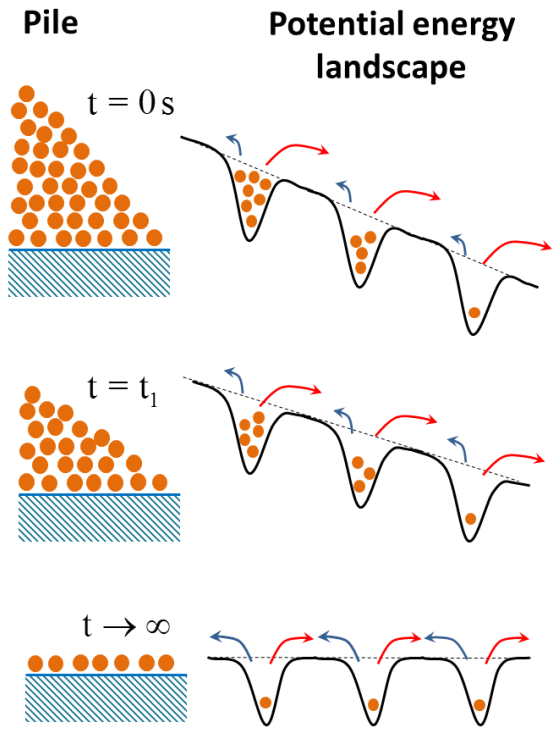

Figure 4. Jumping out of the well. A classical "washboard potential" model useful to understand the relaxation of piles of grains or superconducting vortices.

The overall inclination of the washboard potential is proportional to the difference in the number of grains trapped into adjacent wells. Due to the taps, the grains may eventually escape from the potential wells: they have a large probability to escape downhill (red arrows), and a smaller one to escape uphill (blue arrows).The second row in Fig. 4 is analogous to the first one, but after the application of a certain number of taps. The net downhill motion of grains has reduced the slope of the pile and, correspondingly, the slope of the washboard potential. The smaller inclination implies that the probability to escape downhill is smaller than before, while the probability to escape uphill is larger than before: that implies that the net downhill motion is smaller, and the inclination process is slower. After an infinite number of taps we get to the bottom row in Fig. 4, where the washboard potential has become horizontal, and the probabilities to escape uphill and downhill are identical: there is no net granular motion. So, the system has reached equilibrium, and the washboard potential stops evolving. My description can be put in mathematical terms using the methods of classical statistical mechanics where the taps mimic an "effective temperature", which allows to demonstrate that the slope of the pile decreases logarithmically as the number of taps (i.e., as the time) increases. A similar model can be used for the case of superconducting vortices, resulting in a logarithmic decrease of magnetization [25], and potentially explains a phenomenon that occurs in magnetic materials called "magnetic viscosity" [26].

By the beginning of the 1990's I was familiar with superconducting vortex physics: our group systematically made relaxation experiments, for example. Then, I discovered the BTW model for a pile of grains, which motivated me to search for avalanches in vortex systems. Over the years, I concentrated more and more on avalanches and relaxation of granular matter. All in all, a beautiful analogy between distant fields of Physics has modulated my scientific career for decades. It has been a joyful trip.